\documentstyle[12pt]{article}
\begin{document}

\title{An Example of the Decoherence Approach \\
to Quantum Dissipative Chaos}

\author{Todd A. Brun\thanks{Original address: Physics Department,
California Institute of Technology, Pasadena, CA  91125} \\
Department of Physics, Queen Mary and Westfield College, \\
London  E1~4NS  ENGLAND}

\date{\today}

\maketitle

\begin{abstract}
Quantum chaos---the study of quantized nonintegrable Hamiltonian
systems---is an extremely well-developed and sophisticated field.
By contrast, very little work has been done in looking at quantum
versions of systems which classically exhibit {\it dissipative} chaos.
Using the decoherence formalism of Gell-Mann and Hartle, I find a
quantum mechanical analog of one such system,
the forced damped Duffing oscillator.
I demonstrate the classical limit of the system, and discuss its decoherent
histories.  I show that using decoherent histories, one can define not
only the quantum map of an entire density operator, but can find
an analog to the Poincar\'e map of the individual trajectory.  Finally,
I argue the usefulness
of this model as an example of quantum dissipative chaos, as well as of
a practical application of the decoherence formalism to an interesting
problem.
\end{abstract}

\section{Introduction}

\subsection{Classical laws and dissipative chaos.}

Recently, Gell-Mann and Hartle, among others \cite{GMHart,DowkHall,Brun},
have studied the problem of classical laws arising from quantum theory in the
light of the {\it decoherence formalism}.
In this approach, one considers possible
histories of a given system to which probabilities can be assigned that obey
classical probability sum rules.  In order for histories to {\it decohere}
in this way, it is usually necessary to coarse-grain the description of the
system, by giving the values of its variables only at certain times, or
averaged over certain intervals, or by neglecting certain variables and
retaining others, or a combination of all of these.

They have found that within this formalism it is possible to define in a very
rigorous way the classical equation of motion based only on the underlying
quantum theory.  In doing so, both dissipation and noise typically appear,
arising as a consequence of coarse-graining over neglected degrees of
freedom.  This takes advantage of the well-studied phenomenon of
environmentally induced decoherence \cite{Zurek}.
In addition to casting new light on the problem of how classical
laws of physics arise, this provides an unparalleled tool for studying
quantum mechanical systems with dissipation, and seeing how this alters
their behavior from the more usual Hamiltonian behavior of closed systems.
One area which can profitably be treated this way is dissipative chaos.

There has been an enormous amount of work done on ``Quantum Chaos,''
i.e., quantizing nonintegrable Hamiltonians which classically exhibit
chaotic behavior.  This has turned up beautiful connections between
classical chaotic behavior and their quantum quasiperiodic equivalents.
But very little has been done in looking at the quantum versions of systems
which classically exhibit dissipative chaos,
or on looking at their classical
limit \cite{Graham,ZhengSavage}.  Classical dissipative chaos is
qualitatively very different from Hamiltonian chaos, and one would
expect their quantum equivalents to reflect this difference, but
this has not been widely investigated.
Indeed, even very extensive treatments
of quantum chaos rarely deal with dissipative systems at all
\cite{Gutzwiller}.

There are a number of reasons for this.  The first is that dissipation is
difficult to treat in normal quantum mechanics.
The usual Schr\"odinger equation
is only valid for closed systems without friction.  Open systems in general,
and dissipation in particular, can be handled using the influence functional
approach of Feynman and Vernon \cite{FeynVern}; this has been done in the
case of Brownian motion by Caldeira and Leggett \cite{CaldLegg} among others.
This approach has not been widely used, though, until recently,
as it involves considerable
conceptual and mathematical baggage \cite{Zurek}.  Also, the types of
behavior of most interest to those who study quantum Hamiltonian chaos
involve the coherent evolution of the wave function, with its attendant
complicated structure (e.g., the ``scarring'' of energy eigenfunctions
about classical periodic orbits, the statistics of energy level spacing).
The presence of strong damping wipes out this coherent structure.

Chirikov et al. have summed up the usual attitude towards quantum dissipative
chaos:  ``In what follows we will discuss only Hamiltonian (nondissipative)
systems, considering them to be the more fundamental ones.  Phenomenological
friction is but a crude approximation of the molecular Hamiltonian chaos
which is inevitably related to some noise according to the
fluctuation-dissipation theorem.''  And further, they divide the problem
of quantum chaos into two parts, the quantum dynamics of the wave function
in isolation, and the results of measurement ``with its unavoidable
statistical effect of the irreversible $\psi$ collapse which is a sort of
inevitable noise.'' \cite{Chiretal}

While this is undeniably true, most systems are {\it not} isolated, and
so it is perhaps useful to consider systems for which dissipation is
important. Dissipative chaotic systems may not be fundamental, but they
are nevertheless interesting.
Decoherence is an appropriate formalism in which to
study them \cite{Brun2}.

In the rest of this section I give a brief introduction to the decoherent
histories formalism of Gell-Mann and Hartle.  Then in section 2 I derive
a model for a quantum forced, damped nonlinear oscillator, following
the usual system/environment coarse-graining.  In section 3 I discuss the
classical properties of the forced, damped Duffing oscillator, and
describe some of the properties of dissipative chaos which it exhibits.
In section 4 I treat the quantum version of this problem, and show
how one can make close contact with the classical theory using the
decoherent histories formalism.  In section 5 I illustrate this with a
numerical example, and in section 6 I summarize my conclusions.

\subsection{Decoherent histories.}

In the formalism of decoherent histories, systems are described by a
set of exclusive and exhaustive histories $\{\alpha\}$, which can be
thought of as different possibilities for the system's evolution.  While
there is a vast range of possible sets of histories to choose from, these
sets are restricted by the {\it decoherence condition}
\begin{equation}
D[\alpha,\alpha'] = p_\alpha \delta_{\alpha\alpha'},
\end{equation}
where $D[\alpha,\alpha']$ is the {\it decoherence functional} and
$p_\alpha$ is the probability of history $\alpha$ occurring.  This
decoherence condition restricts one to histories which obey the usual
probability sum rules; these histories do not interfere with each other.

To make this more explicit, in ordinary nonrelativistic quantum mechanics
one can specify a history by enumerating a complete set of orthogonal
projection operators $\{P^i_{\alpha_i}(t_i)\}$ at a sequence of times
$t_i$.  A single history is then given by choosing one projection operator
at each time.  This is equivalent to enumerating a set of possible assertions
about the system at a sequence of times, and having each history be a
string of such assertions.  One can define a {\it history operator}
\begin{equation}
C_\alpha = P^n_{\alpha_n}(t_n)
  \cdots P^2_{\alpha_2}(t_2) P^1_{\alpha_1}(t_1),
\end{equation}
where $\alpha$ is a shorthand for the choices $\alpha_i$ at times $t_i$.
The decoherence functional is then
\begin{equation}
D[\alpha,\alpha'] =
  {\rm Tr} \bigl\{ C_\alpha \rho C^\dagger_{\alpha'} \bigr\}.
\label{DecoherenceFunctional0}
\end{equation}
The density operator $\rho$ is the system's initial condition.

As a rule, it is impossible for very fine-grained histories to decohere;
thus, considerable coarse-graining is required.  One very common
coarse-graining used to study decoherence in systems with many
degrees of freedom is to completely trace out certain freedoms (the
``environment'') while leaving others completely fine-grained (the
``disinguished subsystem'').  This was first studied by Feynman and Vernon
\cite{FeynVern} and applied to decoherence by Zurek \cite{Zurek} among
others.  We will initially be considering this type of coarse-graining.

\section{The Model}

The particular model we will study is based on earlier work on decoherence
in systems with dissipation \cite{Zurek,GMHart,Brun,Brun2}.
In this model we will divide our system into a distinguished variable $x$,
termed the {\it system variable}, and a set of {\it reservoir variables}
$\{ Q_k \}$ which we will trace over.
This system and reservoir will have a total action
\begin{equation}
S[x(t),{\bf Q}(t)] = S_{\rm sys}[x(t)] + S_{\rm res}[{\bf Q}(t)]
  + \int_{t_0}^{t_f} V_{\rm int}(x(t),{\bf Q}(t)) dt,
\end{equation}
where the system variable will be treated as a particle moving in a potential
\begin{equation}
S_{\rm sys}[x(t)] = \int_{t_0}^{t_f} \biggl( {M\over2}{\dot x}^2(t) -
  U(x(t)) \biggr) dt,
\end{equation}
the reservoir is approximated as a collection of harmonic oscillators
\begin{equation}
S_{\rm res}[{\bf Q}(t)] = {m\over2} \sum_k \int_{t_0}^{t_f} \biggl(
  {\dot Q}_k(t)^2 - \omega_k^2 Q_k(t)^2 \biggr) dt,
\end{equation}
and the interaction is linear in $x$ and {\bf Q}:
\begin{equation}
V_{\rm int}(x,{\bf Q}) = - x \sum_k \gamma_k Q_k.
\end{equation}
We will make the additional assumption that the initial density matrix of
the system and reservoir factors, and that the reservoir is initially in a
thermal state.  Then $\rho_{\rm total}(x,{\bf Q}; x^\prime, {\bf Q}^\prime) =
\chi(x; x^\prime) \psi_0({\bf Q}; {\bf Q}^\prime)$, where $\psi_0 = \rho_T$
is a thermal density operator at temperature $T$.

The {\it decoherence functional} in this coarse-graining is then
\begin{equation}
D[x^\prime(t),x(t)] = \exp{i\over\hbar}\biggl\{ S_{\rm sys}[x^\prime(t)]
  - S_{\rm sys}[x(t)] + W[x^\prime(t),x(t)] \biggr\} \chi(x_0; x^\prime_0).
\label{DecoherenceFunctional}
\end{equation}
$W[x^\prime(t),x(t)]$ is the {\it influence phase}, which includes the
collective effects of the traced-over reservoir degrees of freedom.  As
shown by Caldeira and Leggett \cite{CaldLegg}, this functional is
\begin{eqnarray}
&& W[x^\prime(t),x(t)] =  \sum_k {{i\gamma_k^2}\over{m\omega_k}}
  \coth(\hbar\omega_k/kT) \nonumber\\
&& \times \int_{t_0}^{t_f} dt
  \int_{t_0}^{t_f} ds\ \cos(\omega_k(t-s)) (x^\prime(t) - x(t))
  (x^\prime(s) - x(s)) \nonumber\\
&& - {{\gamma_k^2}\over{2m\omega_k}} \int_{t_0}^{t_f} dt \int_{t_0}^t ds\
  \sin(\omega_k(t-s)) (x^\prime(t) - x(t)) (x^\prime(s) + x(s)).
\end{eqnarray}

We can now switch to new variables
$X = (x + x^\prime)/2$ and $\xi = x - x^\prime$.
In these variables
\begin{eqnarray}
&& S_{\rm sys}[x^\prime(t)] - S_{\rm sys}[x(t)] = \nonumber\\
&& \int_{t_0}^{t_f} \xi(t) \biggl(
  - M{\ddot X}(t) - {{dU}\over{dt}}(X(t)) \biggr) dt
  - \xi_0 M{\dot X}_0 + O(\xi^3),
\end{eqnarray}
which as we see contains the Euler-Lagrange equations.  We also go to a
continuum of oscillator frequencies with a Debye distribution, in which the
discrete sums become integrals over a weighting function $g(\omega) =
\eta \omega^2 \exp(-\omega/\Omega)$, where $\Omega$ is a large cutoff
frequency, so that $1/\Omega << t_f - t_0$.  In this limit,
the influence phase becomes
\begin{eqnarray}
&& W[X(t),\xi(t)] = \int_{t_0}^{t_f} \xi(t) \biggl( - M \Lambda X(t)
  - 2 \Gamma {\dot X}(t) \biggr) dt \nonumber\\
&& + i K \int_{t_0}^{t_f} \xi^2(t) dt + O(\Omega^{-2}).
\end{eqnarray}
where $\Lambda = \eta\Omega/m$, $\Gamma = \pi\eta/4mM$,
and $K = 4M\Gamma kT/\hbar$.  The
$\Lambda$ term has the form of a linear force; it can be absorbed
into the system action by going to an effective potential
\begin{equation}
U_{\rm eff}(X) = U(X) + M \Lambda X^2/2.
\end{equation}
The $\Gamma$ term has the form of a dissipation.

The imaginary term is of particular interest.
It suppresses $D[X(t),\xi(t)]$ when $\xi \ne 0$.  Since $\xi \ne 0$
corresponds to the ``off-diagonal'' terms of the decoherence functional
$(x(t) \ne x^\prime(t))$, the suppression of these terms results in
approximate decoherence of this set of histories.
This suppression of off-diagonal terms is
clearly related to the presence of noise \cite{Zurek,GMHart}.
The kernel of this term can be identified with
the two-time correlation function of
a stochastic driving force $F(t)$ in the
quasiclassical limit.
This correlation function is
\begin{equation}
\langle F(t) F(s) \rangle = \hbar K \delta(t-s)
\end{equation}
in the continuum case, with $\langle F(t) \rangle = 0$.  So in the
quasiclassical limit this system obeys the classical equation of motion
\begin{equation}
{\ddot x} + {1\over M}{dU_{\rm eff}\over{dx}}(x) + 2\Gamma{\dot x}
  = F(t)/M.
\label{EOM1}
\end{equation}

Instead of taking the reservoir to be in a thermal state initially, we
can take it to be in a {\it displaced} thermal state,
\begin{equation}
\rho_{\rm DT} = {\hat D}(q(\omega),p(\omega)) \rho_T
  {\hat D}(q(\omega),p(\omega))^\dagger,
\end{equation}
${\hat D}(q(\omega),p(\omega))$ is the
coherent state displacement operator:
\[
 {\hat D}(q(\omega),p(\omega)) | 0 \rangle = | q(\omega),p(\omega) \rangle,
\]
where $| q(\omega),p(\omega) \rangle$ is the coherent state centered on
$(q(\omega),p(\omega))$ at frequency $\omega$.
If we take $q(\omega)$ to be sharply peaked around a certain
frequency, $q(\omega) = q \delta(\omega - \omega_0)$,
and $p(\omega) = 0$, then in the
quasiclassical limit the above equation of motion (\ref{EOM1}) gains an
additional term
\begin{equation}
{\ddot x} + {1\over M}{dU_{\rm eff}\over{dx}}(x) + 2\Gamma{\dot x}
  = q \cos(\omega_0 t) + F(t)/M.
\label{EOM2}
\end{equation}
This is exactly the form for a nonlinear oscillator with damping and a
periodic driving force, with additional noise.

In a truly classical system,
$F(t)$ would vanish as $T\rightarrow0$,
but in the quantum theory noise is always present, even at absolute zero.
One can think of it as arising from the
zero-point oscillations of the reservoir oscillators.  At low temperatures,
however, the two-time correlation function of the noise is
highly {\it nonlocal in time}.  At $T = 0$,
\begin{equation}
{\rm Re}\ W[X(t),\xi(t)] \sim \int_{t_0}^{t_f} dt \int_{t_0}^{t_f} ds\
  \xi(t)\xi(s)/(t-s)^2.
\end{equation}
Correlations in the noise persist for all times.  Because of this form of the
kernel, doing exact (or even numerical) calculations in the low-temperature
limit is extremely difficult.  This is why the high $T$ limit is generally
used.

\section{The Classical Forced Damped Duffing Oscillator}

We are interested in finding quantum equivalents to classical systems
which exhibit dissipative chaos.  While many such systems (e.g., fluid
mechanics) have no easily realizable quantum limit, there are some
which can be readily quantized as shown in section 2.  These are the
nonlinear oscillators with damping and driving.

One much-studied classical nonlinear oscillator is the Duffing oscillator,
characterized by a two-welled polynomial potential,
\begin{equation}
U(x) = {x^4\over4} - {x^2\over2}.
\end{equation}
With forcing and damping, this gives an equation of motion
\begin{equation}
{\ddot x} + (x - x^3) + 2\Gamma{\dot x}
  = q \cos(\omega_0 t),
\label{EOM3}
\end{equation}
where we take $M=1$.  This system is chaotic for certain values of $q$,
$\Gamma$, and $\omega_0$ \cite{GuckHolmes}.  For example, $q=0.3$,
$\Gamma=0.125$ and $\omega_0 = 1.0$ is a common choice.

Since this system has explicit time dependence,
its phase space is three-dimensional, $(x,p,t)$.  It is
common to discretize the dynamics by taking a {\it Poincar\'e section},
considering only the points on a surface of constant phase $(x_i, p_i)$ at
times $t_i = 2\pi i/\omega_0$.  The continuous dynamics defines a {\it map}
${\bf f}$:
\begin{equation}
(x_i, p_i) \rightarrow (x_{i+1}, p_{i+1}) = {\bf f} (x_i, p_i).
\label{poincare_map}
\end{equation}
If the oscillator is non-chaotic, there is a stable attracting fixed point
or group of periodic points to which the $(x_i, p_i)$ quickly tend.
These points
correspond to a periodic orbit of the continuous dynamics.  When the
oscillator becomes chaotic, the stable set becomes a {\it strange
attractor}, a fractal structure with non-periodic behavior.  There are, in
addition, an infinite number of unstable fixed points and periodic points.
(See figure 1.)

We can also look at the classical dynamics from the point of view of
probability measures $P(x,p)$ on phase space.  The map ${\bf f}$ of
points in phase space induces a map on probability measures
\begin{equation}
P_i(x,p) \rightarrow P_{i+1}(x,p) = \int dx^\prime dp^\prime\
  \delta( (x,p) - {\bf f}(x^\prime,p^\prime) ) P_i(x^\prime, p^\prime).
\label{ProbabilityMeasure}
\end{equation}
By means of this sort of map we can readily make contact with the quantum
theory.

Of particular interest are
{\it invariant} probability measures $P_{\rm inv}$.
There are many of these, most corresponding to unstable fixed points and
periodic points of the map ${\bf f}$.  It is possible to eliminate these
unstable solutions by including a small amount of noise in the equation of
motion (\ref{EOM3}).  This effectively broadens the delta functions in
(\ref{ProbabilityMeasure}) into peaks of finite width $\epsilon$, and
eliminates all unstable solutions, leaving a single unique $P_{\rm inv}$
corresponding to the strange attractor.  Classically, we can then allow
the noise to go to zero, and look at $P_{\rm inv}$ in the zero-noise
limit.  In that limit, the invariant probability measure becomes a
generalized function with substructure at all length scales.
It is a fractal.

\section{Decoherent Histories, Quantum Maps, and Probability}

At best, the functional described in (\ref{DecoherenceFunctional}) can only
be approximately decoherent.  Clearly, off-diagonal terms will not vanish
for sufficiently small $|\xi(t)|$.  More coarse-graining is needed in the
description of $x(t)$ and $x^\prime(t)$.  Also, specifying a value, even an
approximate value, of $x(t)$ for all times $t$ is an extreme fine graining.
It is more common to instead specify $x$ at a series of discrete times $t_i$.
Thus, instead of a complete trajectory $x(t)$ one gives only a series of
$x$ values $\{x_i\}$.  Coarse-graining in position as well, one could divide
up the range of $x$ into finite non-zero intervals $\Delta_j^i$,
where $j$ is an index specifying which interval $x$ fell in at time $t_i$.
A history would now be a series of indices $\{\alpha_i\}$, specifying that
$x$ fell in the interval $\Delta_{\alpha_i}^i$ at time $t_i$.
Note that to achieve decoherence, these times $t_i$ cannot be too close
together; they must generally be separated by at least the {\it
decoherence time} \cite{GMHart}.  For high temperature systems this
is typically quite short, of the order $\hbar^2/2M\Gamma kT d^2$,
where $d$ is the size of the intervals.

Such a coarse-graining gives us a new decoherence functional:
\begin{equation}
D[\alpha,\alpha^\prime] = \int_\alpha \delta x
  \int_{\alpha^\prime} \delta x^\prime \
  D[x(t),x^\prime(t)],
\label{DecoherenceFunctional2}
\end{equation}
where the limits specify integration only over those paths which pass
through the series of intervals $\Delta^i_{\alpha_i}$ at the times $t_i$.
The probability of a given history $\alpha$ is of course given by the
diagonal terms of this functional.  Since the original decoherence functional
given by (\ref{DecoherenceFunctional}) has an exponent quadratic in $\xi$,
the path integrals over $\xi$ can be carried out; we then let $\alpha =
\alpha^\prime$ and get
\begin{equation}
p(\alpha) = \sqrt{2\pi\over K} \int_\alpha \delta X\ \exp\biggl\{
  - {1\over{K\hbar}} \int_{t_0} e^2(t) dt \biggr\} w(X_0,M{\dot X}_0 ),
\label{Probability}
\end{equation}
where $e(t) = M {\ddot X} + (dU_{\rm eff}/dt)(X) + 2M\Gamma{\dot X}
-q\cos(\omega_0 t)$ is the right-hand side of the equation of motion
\cite{Brun2}.
{}From this we see that the probability will by peaked about histories which
approximately obey the classical equation of motion  $e(t) = 0$, more and
more sharply as we approach the classical limit where $M$ is large.
The $w(X_0, M{\dot X}_0)$ is the initial Wigner distribution of the system.

The Wigner distribution is defined in terms of the density matrix:
\begin{equation}
w(X,p) = {1\over\pi} \int e^{- i\xi p/\hbar} \chi(X+\xi/2; X-\xi/2)\ d\xi.
\end{equation}
The distribution behaves very
similarly to a classical probability distribution
on phase space, except that $w(X,p)$ can be locally negative (though it must
sum to 1 over all of phase space, and be non-negative on average over regions
with volumes larger than $\hbar$).  The expectation values of functions of
$X$ and $p$ can be calculated by averaging them over phase space using
$w(X,p)$ as a weighting function,
though there is usually some ambiguity about
the ordering of operators.  As one goes to the classical limit, on scales
large compared to $\hbar$, this ambiguity becomes unimportant.

An interesting way of looking at this system is in terms of the evolution of
the Wigner distribution with time.
If we consider surfaces of constant phase,
as in the classical case, we can define a {\it quantum map},
\begin{eqnarray}
&& w_i \rightarrow w_{i+1} = {\bf T} w_i, \nonumber\\
&& w_{i+1}(X_1,p_1) = \int dX_0 \int dp_0\ T(X_1,p_1; X_0, p_0) w(X_0,p_0).
\end{eqnarray}
The transition matrix ${\bf T}$ is defined by the path integral
\begin{eqnarray}
&& T(X_1,p_1; X_0,p_0) = {1\over\pi} \int d\xi_0 d\xi_1
  e^{ -i (\xi_1 p_1 - \xi_0 p_0) } \nonumber\\
&& \times \int \delta X \delta \xi\
  \exp {i\over\hbar} \biggl\{ S_{\rm sys}[X(t) + \xi(t)/2] \nonumber\\
&& - S_{\rm sys}[X(t) - \xi(t)/2] + W[X(t),\xi(t)] \biggr\},
\end{eqnarray}
\begin{eqnarray}
&& = {1\over\pi} \int d\xi_0 d\xi_1
  e^{ -i (\xi_1 p_1 - \xi_0 p_0) } \nonumber\\
&& \times \int \delta X\
  \exp \biggl\{ - {1\over{\hbar K}} \int_{t_i}^{t_{i+1}} e^2(t) dt
  + i ( \xi_1 M {\dot X}_1 - \xi_0 M {\dot X}_0 ) \biggr\}, \\
&& = 4 \pi \int \delta X\ \delta(p_0 - M{\dot X}_0) \delta(p_1 - M{\dot X}_1)
  \exp \biggl\{ - {1\over{\hbar K}}
  \int_0^{2\pi/\omega_0} e^2(t) dt \biggr\}.
\nonumber
\label{TransitionMatrix}
\end{eqnarray}
This evolution strongly resembles the classical evolution of probability
measures induced by the phase-space map, and in the classical limit we expect
$w(X,p)$ to evolve towards an invariant distribution $w_{\rm inv}(X,p)$ which
closely resembles the classical invariant measure $P_{\rm inv}(x,p)$.
Graham \cite{Graham} has demonstrated this sort of behavior in his work
on the quantum Lorenz model, which, though very different in approach
from this paper, may nevertheless be indicative;
and the author's own numerical
simulations \cite{Brun3} seem to bear this out
(though of course one would not
expect numerical simulations to exhibit unstable alternative solutions).

In the quantum case,
it is impossible for the noise to ever truly vanish.  Even
at absolute zero, zero-point fluctuations remain that prevent $w(X,p)$ from
becoming a true fractal.  Though the invariant distribution may strongly
resemble $P_{\rm inv}$ for a wide range of scales, there is always some
scale at which the quantum noise ``smears out'' $w_{\rm inv}(X,p)$.

While these maps on Wigner distributions make contact with the classical
theory, ideally we would like to find some quantum analog to
(\ref{poincare_map}), i.e., a description in terms of individual histories,
rather than probability distributions.  To do this, let us consider
yet another coarse graining.  Consider the decoherence functional
(\ref{DecoherenceFunctional0}), where we take the sequence of times
to be those corresponding to the surface of section $t_i = 2\pi i$,
and let the projection operators $P_{q,p}$ be onto localized cells
of phase space centered at $(q,p)$.  While there are no true projections
onto cells of phase space, there are approximate projectors which
can be used to get approximate decoherence \cite{Halliwell,Brun2}.
For example, simple coherent state projections $|q,p\rangle\langle q,p|$
can be used.

A history can then be specified by a series of points $\{q_i,p_i\}$ at
the times $t_i$, and the decoherence functional calculated
\begin{equation}
D[\{q_i,p_i\},\{q_i',p_i'\}] =
  {\rm Tr} \bigl\{ P_{q_n,p_n} {\bf T}( \cdots
  {\bf T}(P_{q_1,p_1} \rho_0 P_{q_1',p_1'} \cdots ) P_{q_n',p_n'} \bigr\}.
\end{equation}
Here we have taken {\bf T} to be the transition matrix on density operators
rather than Wigner distributions; it is simple to go from one representation
to the other.  This is a {\it quantum surface of section}.  At each time
$t_i$ the system is localized in a cell in phase space centered on
$(q_i,p_i)$, and probabilities can be assigned to each possible next point
$(q_{i+1},p_{i+1})$.  This differs, of course, from the classical case
where the evolution is deterministic; but from (\ref{Probability}) it
is clear that these histories will be peaked about the classical evolution
in the quasiclassical limit.
This is shown explicitly by the numerical example in the next section.

\section{Numerical simulation and quantum state diffusion}

While the decoherent histories formalism has great interpretational
power, it is not very convenient for numerical simulation.  Enumerating
all the possible histories and calculating the elements of the decoherence
functional is a daunting task.  What one would like is a method of
generating the histories with the correct probabilities without having
to solve the full master equation at every step.

Recently, it has been shown that the theory of quantum state diffusion
provides just such a technique.  Quantum state diffusion is one of
the so-called {\it quantum trajectory} methods, in which the master
equation evolution is unravelled into quantum trajectories
of individual states.  These states obey a nonlinear stochastic differential
equation which in the mean reproduces the master equation.  Because
one need deal only with a single quantum state at a time, this is
very suitable for numerical calculation \cite{QSD}.

Diosi, Gisin, Halliwell and Percival \cite{DGHP}
have shown that these individual
quantum trajectories correspond to a set of approximately decoherent
histories.  In the case of a dissipative interaction, these correspond
to histories of systems localized into small cells in phase space,
and their probabilities match those given by decoherent histories.
Thus it is ideal for the sort of problem we are interested in.

For further details see the references.  More work on the connections
between decoherent histories and quantum state diffusion, and their
application to dissipative chaos, is currently underway \cite{Brun3}.

In figure 2 we see one such trajectory, generated in the quasiclassical
limit (where $\hbar = 10^{-4}$).  One can see that this is very
close to the classical limit, but with additional ``smearing'' due to
the presence of noise.  This smearing sets a lower cutoff scale to the
substructure of the strange attractor.  As one continues to go to the
classical limit, more and more substructure appears, and the noise
becomes less and less important.

Note that in this chaotic system we expect the trajectory to evenly
sample the ``invariant'' Wigner distribution $w_{\rm inv}$ over time.
{}From the distribution of points we see that this is indeed very close
to the structure of the classical strange attractor in figure 1.

\section{Conclusions}

As we have seen, it is possible to use the decoherence formalism to study
at least some systems exhibiting classically chaotic behavior, and to
do so in a way which includes dissipation in a simple and natural fashion.
Though the model treated here is very much a special case, intended only
to illustrate the basic ideas of the theory, it is remarkable how many
details can be brought out and studied by its means.

Certainly these techniques should work for any kind of nonlinear oscillator,
or for multivariable extensions of them.  It might well be possible to
treat systems of experimental interest, arising in fields such as quantum
optics.  Some work on such systems has already been done by other workers
\cite{Savage}.

Using the usual master equation formalism it is possible to draw a
close connection between the classical theory of probability measures
and the quantum Wigner distribution.  But with decoherent histories,
one can also find a quantum analog to individual chaotic orbits,
such as the {\it quantum surface of section} defined in section 4.

One can then argue analytically that these quantum histories become
more and more closely peaked about the classical equations of motion
as one goes to the classical limit; and this correspondence can also
be demonstrated numerically.

Further analytical
study may yield better results for the probabilities and decoherence of
phase space histories.  And it may be fruitful to explore what equivalents
there are in the quantum case to classical quantities such as Lyapunov
exponents, fractal dimension, and Kolmogorov entropy.
This theory should amply reward further study, both analytical
and numerical.

\section*{Acknowledgements}

First and foremost I would like to acknowledge the guidance of my
advisor, Murray Gell-Mann, and many stimulating discussions with
my Seth Lloyd, Juan Pablo Paz, Wojciech Zurek, and Jim Hartle.  Considerable
help on phase space histories was provided by Jonathan Halliwell, and
I am indebted to Ian Percival and Nicolas Gisin for my knowledge of
quantum state diffusion theory.  This work was supported in part by
DOE Grant No. DE-FG03-92-ER40701.

\vfil\eject

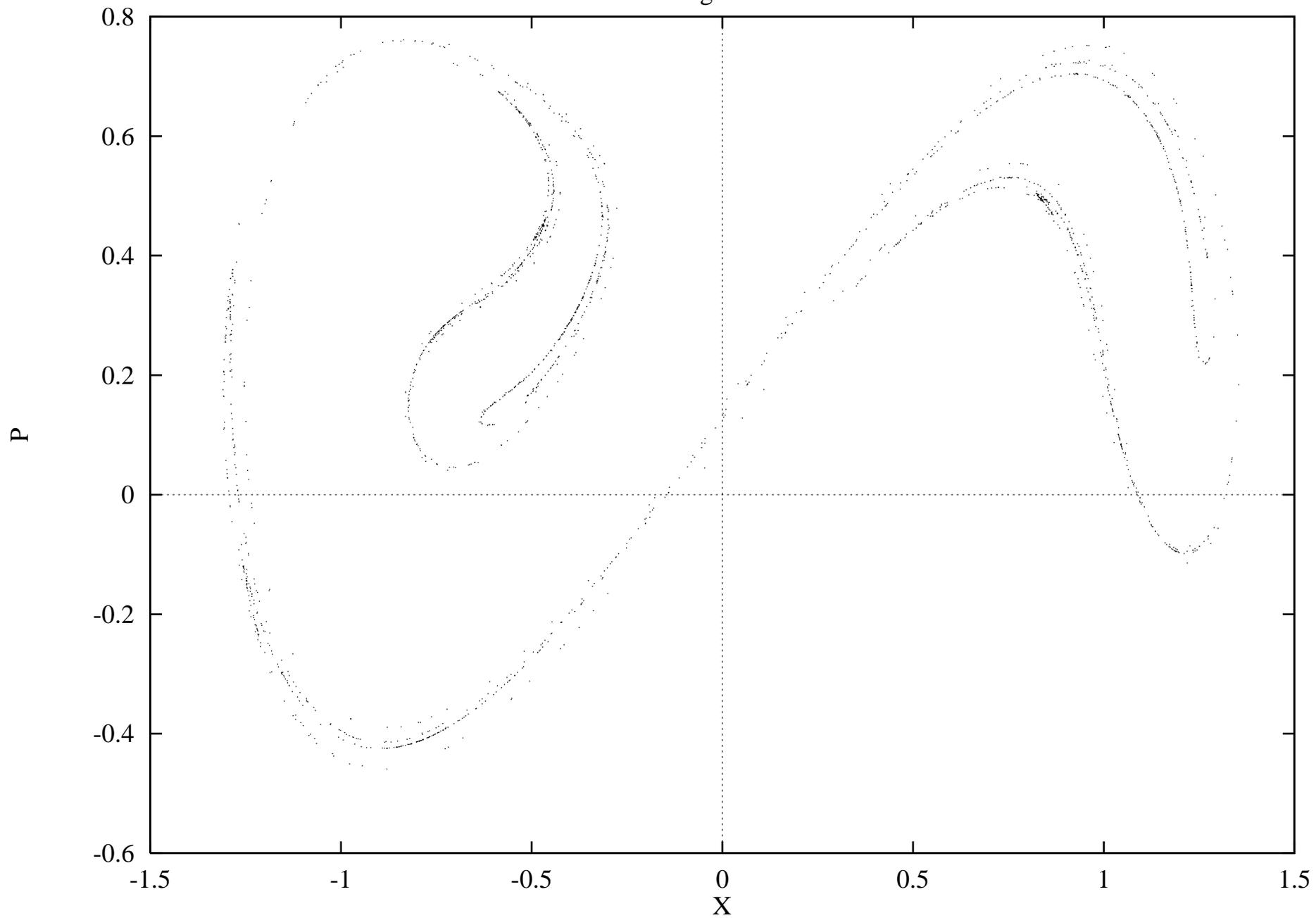
Figure 1.  The classical forced damped Duffing oscillator surface of
section in the chaotic regime.
$q = 0.3$, $\Gamma = 0.125$, $\omega_0 = 1.0$.
\vfil

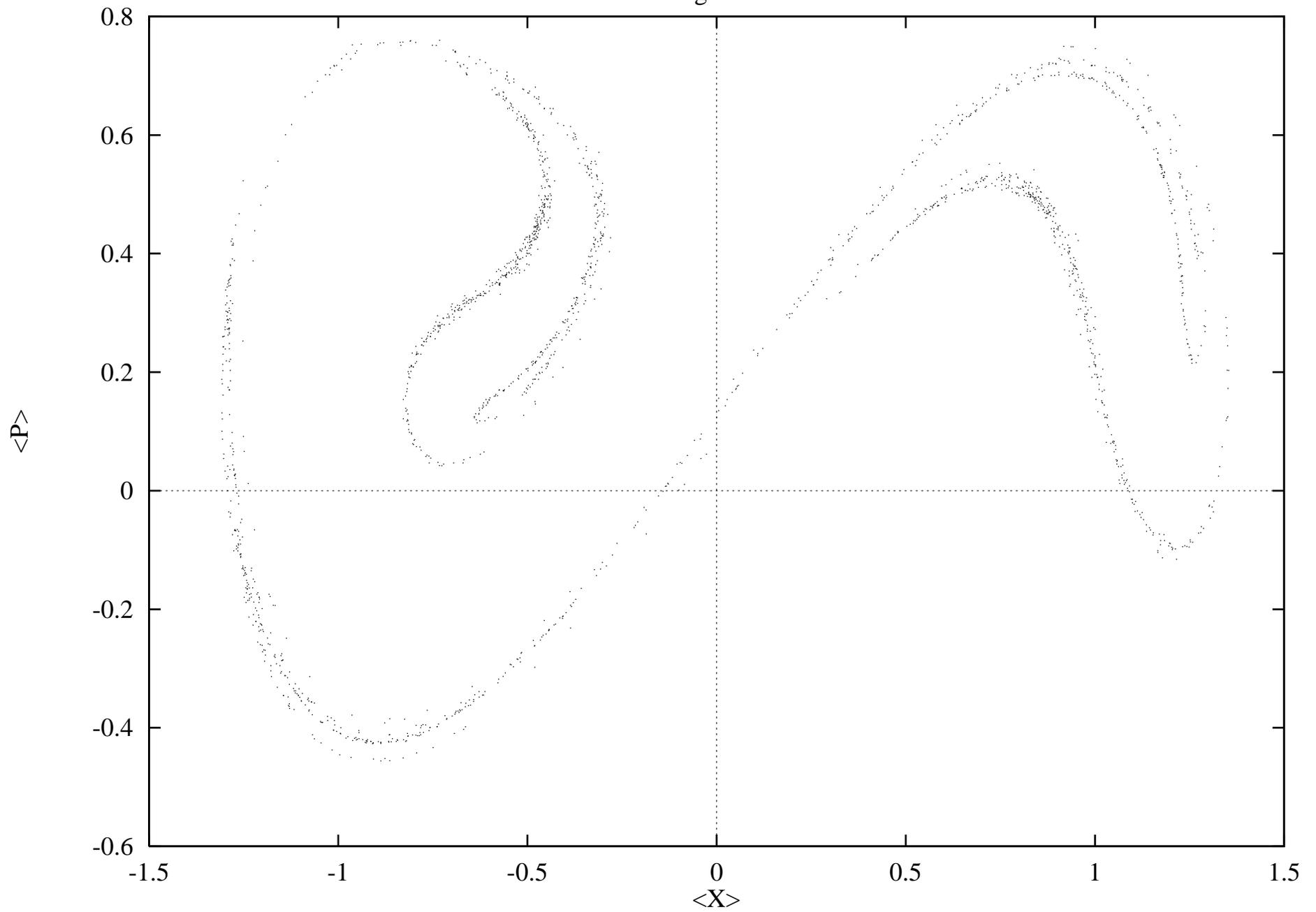
Figure 2.  The quantum forced damped Duffing oscillator surface of
section, generated by the quantum state diffusion algorithm in the
quasiclassical limit.
$\hbar = 10^{-4}$, $q = 0.3$, $\Gamma = 0.125$, $\omega_0 = 1.0$.
\vfil

\end{document}